\documentclass[sigconf]{acmart}
\usepackage{framed}
\usepackage{listings}
\AtBeginDocument{%
  }

\copyrightyear{2026}
\acmYear{2026}
\setcopyright{cc}
\setcctype{by}
\acmConference[ICPC '26]{34th IEEE/ACM International Conference on Program Comprehension}{April 12--13, 2026}{Rio de Janeiro, Brazil}
\acmBooktitle{34th IEEE/ACM International Conference on Program Comprehension (ICPC '26), April 12--13, 2026, Rio de Janeiro, Brazil}
\acmPrice{}
\acmDOI{10.1145/3794763.3794829}
\acmISBN{979-8-4007-2482-4/2026/04}

\begin{document}

\title{CMind: An AI Agent for Localizing C Memory Bugs}

\author{Chia-Yi Su}
\email{csu3@nd.edu}
\affiliation{%
  \institution{University of Notre Dame}
  \city{Notre Dame}
  \state{Indiana}
  \country{USA}
}

\author{Collin McMillan}
\email{cmc@nd.edu}
\affiliation{%
  \institution{University of Notre Dame}
  \city{Notre Dame}
  \state{Indiana}
  \country{USA}
}

\renewcommand{\shortauthors}{Su et al.}

\begin{abstract}
This demonstration paper presents CMind, an artificial intelligence agent for localizing C memory bugs.  The novel aspect to CMind is that it follows steps that we observed human programmers perform during empirical study of those programmers finding memory bugs in C programs.  The input to the tool is a C program's source code and a bug report describing the problem.  The output is the tool's hypothesis about the reason for the bug and its location.  CMind reads the bug report to find potential entry points to the program, then navigates the program's source code, analyzes that source code, and generates a hypothesis location and rationale that fit a template.  The tool combines large language model reasoning with guided decision making we encoded to mimic human behavior.  The video demonstration is available at~\url{https://youtu.be/_vVd0LRvVHI}. 
\end{abstract}

\begin{CCSXML}
<ccs2012>
   <concept>
       <concept_id>10010147.10010178</concept_id>
       <concept_desc>Computing methodologies~Artificial intelligence</concept_desc>
       <concept_significance>500</concept_significance>
       </concept>
   <concept>
       <concept_id>10011007.10011006</concept_id>
       <concept_desc>Software and its engineering~Software notations and tools</concept_desc>
       <concept_significance>500</concept_significance>
       </concept>
 </ccs2012>
\end{CCSXML}

\ccsdesc[500]{Computing methodologies~Artificial intelligence}
\ccsdesc[500]{Software and its engineering~Software notations and tools}
%\ccsdesc[100]{Do Not Use This Code~Generate the Correct Terms for Your Paper}

\keywords{Bug localization, Agentic AI, LLMs, Programmer Attention}

\maketitle

\section{Introduction}

Memory bugs are among the top concerns for code quality in C programs.  Cotroneo~\emph{et al.}~\cite{cotroneo2016bugs} state that ``use of uninitialized data, buffer overflows, and memory leaks'' are the preponderant types of bugs faced by C programmers -- all three are related to memory management.  At the same time memory bugs are among the most difficult to solve.  Memory bugs can manifest as gradual performance degradation detected only in long running programs or as seemingly random crashes not clearly connected to a user input, and there is a tendency for repairs to only partially fix the problem or even introduce new memory bugs~\cite{yin2011fixes}.

%Lorem ipsum dolor sit amet, consectetur adipiscing elit. Nunc vulputate venenatis mauris, eu iaculis risus rhoncus non. Aliquam erat volutpat. Nulla dapibus odio neque, quis auctor nisi auctor a. Fusce eget dictum ex. Nam id vestibulum ligula. Proin libero ex, finibus at est at, vestibulum bibendum ante. Sed quis ligula in mauris tristique sollicitudin. Nunc ex dolor, aliquam vel diam accumsan, molestie faucibus nunc. Phasellus quis odio at ex condimentum vestibulum. Aliquam dictum sapien at felis sagittis venenatis.

Automatic detection and repair of C memory bugs has long been categorized under automatic program repair generally, with approaches ranging from custom neural models trained on examples of buggy code, to prompts containing sections of source code input to pretrained large language models (LLMs), to fine-tuned LLMs, and most recently tool-augmented agents~\cite{chen2025learning, dikici2025advancements, yang2025survey}.   All hold promise, yet as Melo~\cite{melo2025opportunities} points out, even with state-of-the-art LLMs ``achieving reliable integration and maintaining code quality present significant technical and organizational challenges.''  The reason, as Karpathy~\cite{karpathy2024software} describes, is essentially that LLMs have nearly unlimited range in possible answers and are likely to move off track quickly -- and the answer is to design agents with ``partial autonomy'' that ``keeps LLMs on a leash.''

%Quisque id vehicula risus. Mauris commodo velit a ligula aliquet hendrerit. Pellentesque mattis ipsum eget metus tempus scelerisque. In hendrerit enim vel urna condimentum, a maximus elit accumsan. Nunc molestie hendrerit risus, et tempor ligula aliquet id. Suspendisse ac accumsan ipsum. Class aptent taciti sociosqu ad litora torquent per conubia nostra, per inceptos himenaeos. Maecenas maximus aliquet orci sed blandit. Nulla enim dui, dapibus eget nisi iaculis, efficitur auctor dui. Sed tempor odio nec ipsum laoreet, non tincidunt dui molestie. Aliquam dignissim dolor quis lorem consequat, eu aliquet dolor bibendum. Praesent id commodo dui, sit amet ullamcorper nunc. Phasellus id ligula nec turpis sagittis malesuada at at tellus. Etiam id nisl dolor. Etiam scelerisque dignissim ornare. Nunc porttitor aliquet purus. Orci varius natoque penatibus et magnis dis parturient montes, nascetur ridiculus mus. Duis consectetur finibus nibh. Duis ut mi quis enim sollicitudin luctus.

This demonstration paper introduces CMind, a tool-augmented AI agent for localizing C memory bugs.  CMind is a partially autonomous systems that combines LLM reasoning with guided decision making.  By ``LLM reasoning'' we mean that CMind uses an LLM to read and interpret code sections that we present to it.  By ``guided decision making'' we mean that CMind presents the LLM with limited choices and provides access to specific tools' output that we have evidence is useful in making those choices.  A key novel aspect to CMind is we designed the ``guided'' component to mimic human behavior that was observed in empirical studies of programmers who were localizing C memory bugs.  CMind is partially autonomous in that it uses recent LLM tools but keeps the LLM components ``on a leash'' by following patterns of human behavior for the specific task of localizing C memory bugs.

%Vestibulum dignissim ut ipsum in finibus. Integer aliquet eleifend quam eget fermentum. Phasellus ut sapien eget mauris accumsan fermentum. In congue purus auctor lectus feugiat, nec fermentum quam volutpat. Pellentesque mollis pellentesque ante, aliquam scelerisque risus consectetur nec. Cras eu tempus mi. Sed laoreet lectus sit amet imperdiet sollicitudin. Donec porta, quam ac ultricies faucibus, massa eros convallis lectus, id fringilla neque purus non lectus. Aliquam dapibus libero in metus consequat, vel faucibus quam laoreet. Nam ac placerat tortor. In quis lorem orci. Vestibulum pellentesque finibus lorem, id vulputate enim. In interdum, libero ac auctor pretium, leo ligula sollicitudin dolor, eu faucibus arcu tortor a leo. Nullam urna arcu, malesuada sed tellus a, tincidunt auctor dolor. Cras iaculis metus eu metus tempus bibendum.

The process we found that humans follow to find memory bugs essentially breaks into three steps~\cite{smith2025human}: 1) find entry points to the program relevant to the bug by reading the bug report, 2) follow code structure to find other relevant code based on the entry points, and 3) read code statements along the chain of relevant code structure to determine relevance to the bug report.  We designed CMind to follow this same process but use an LLM at points requiring subjective decision making.  Like a person, CMind reads bug reports, finds entry points, and explores programs based on those entry points.  Then it generates a hypothesis about code relevance to the bug and describes that hypothesis in a template we provide.

%Curabitur auctor sem a lobortis viverra. Suspendisse porttitor nunc risus, ac faucibus quam venenatis eu. Maecenas eu magna fringilla, lobortis mauris id, iaculis magna. Phasellus commodo faucibus consectetur. Duis ac magna eget metus auctor bibendum at quis purus. Suspendisse eget fermentum odio. Nam quis rhoncus massa. Praesent porttitor mattis risus, sagittis gravida mi. Donec egestas pulvinar justo, non sollicitudin metus congue vel. Suspendisse ornare a sem in aliquam. Integer tincidunt massa sed nulla lacinia, non varius mauris interdum. Donec tristique diam ac tortor congue, vitae mattis eros consequat. Aliquam eu convallis quam. Integer blandit tortor eu felis porttitor pretium. 

We make CMind available as a command-line Unix tool and via a web platform.  By default CMind uses GPT-o4 for LLM interaction but this is configurable and in principle any LLM can be used.  For the web platform, the user needs only to upload an archive snapshot of his or her code and the text of a bug report.  For the command-line tool, the user needs only to point CMind to the source code folder (or archive file) and a text file containing the bug report.  In a pilot evaluation we demonstrate how to use the tool on twenty actual memory bugs in C programs we found from public repositories.  %CMind is open-source and available via an online appendix\footnote{\url{https://github.com/apcl-research/CMIND}}:

\section{Background/Related Work}
%Table~\ref{tab:blrelated} shows related works for bug localization over several years. 
The directions for bug localization fall into several categories: 1) spectrum-based fault localization 2) model-based fault localization 3) information-retrieval based 4) neural model-based 5) bio-based 6) LLMs-based, and 7) agentic based.

Spectrum-based methods use run-time behaviors to localize bugs. For example, Jones~\emph{et al.}~\cite{jones2005empirical} proposed Tarantula that uses the coverage of the test data for bug localization.  Model-based approaches use probabilistic models. Baah~\emph{et al.} uses probabilistic graphical models to find bugs in the program. Information retrieval-based methods use static information for bug localization. Mahmud~\emph{et al.}~\cite{mahmud2024using} proposed an information retrieval approach that retrieves files with bug based on GUI information. Zhou~\emph{et al.}~\cite{zhou2012where} used bug reports to find the buggy code. Neural-based methods mean localizing the bugs with neural network. Qian~\emph{et al.}~\cite{qian2023gnet4fl} used graph convolution neural network for bug localization. Bio-based methods use clues from human behavior for bug localization such as the work by Muller~\emph{et al.}~\cite{muller2016using} that uses bio-metrics for bug localization. More recently, several LLMs and agentic-based approaches have been proposed. Stein~\emph{et al.}~\cite{stein2025where} used LLMs for bug localization by exploring the attention in LLMs. Chang~\emph{et al.}~\cite{chang2025bridging} proposed LLM-based hierarchical framework for bug localization.  Due to page limit, several other papers along these lines are highlighted in different survey papers~\cite{niu2025when, wong2016survey}.

Although LLMs have shown success in bug localization~\cite{stein2025where, dai2025less, chang2025bridging, liu2024marscode, qin2024agentfl}, LLMs still struggle with hallucinations~\cite{li2025llmbl,zhang2025llm} and understanding long contexts~\cite{kuratov2024babilong, li2024long}. To limit the range of possible answers, we developed our method based on a human study with human programmers~\cite{smith2025human}.  Specifically, we extracted the information that programmers used with tools during the study and limited LLMs to only look at the information used by the programmers. Compared with other agentic-based methods such as~\cite{qin2024agentfl}, the key novelty of this paper is that we design our prompt based on a human study with human programmers.

\section{Approach}
\begin{figure}[h]
    \centering
    \includegraphics[width=0.8\linewidth]{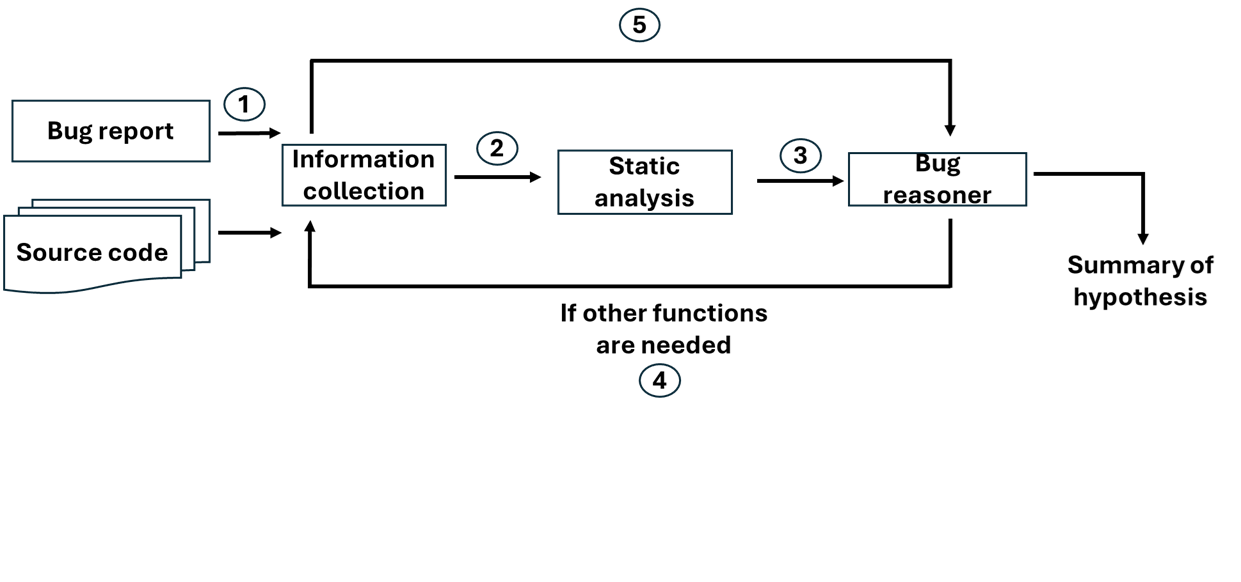}
    \vspace{-10mm}
    \caption{Overview of CMind.  Area 1 is collects the entry point and related functions.  Area 2 generates the static analysis information. Area 3 reasons the bugs based on all given information. Area 4 requests more information. Area 5 returns the requested information to the bug reasoner. }
    \label{fig:overview}
    \vspace{-5mm}
\end{figure}

Figure~\ref{fig:overview} shows the process for CMind. The input of CMind is a bug report and source code and the output is the summary of the hypothesis. Area 1 collects the necessary information including the entry points and related methods based on the bug report. We first used LLMs to determine the function name. Then, we used the Python script to extract the function based on the provided function name or file. Area 2 uses LLMs to determine the type of static analysis to use, i.e. callgraph analysis or dataflow analysis. Then, we used tools to generate those results. Area 3 is the bug reasoner that uses all this information to localize bugs. If the provided information is sufficient for bug localization, the bug reasoner generates hypothesis about the location of the bugs. Otherwise, it would request more information in Area 4 and 5. We design the final prompt based on an empirical study and our experiments. We leave the robustness study as our future work.

%\vspace{-5mm}

\subsection{Information Collection}
Area 1 in Figure~\ref{fig:overview} collects the necessary information. The input is the bug report and the source code. The output is the related functions or source files. We use LLMs to find the program entry points based on the bug report. The entry point that LLMs output could be functions names or files related to the bug report. We limit LLMs to find up to three related functions or files (including the entry function/file) at this time. We limit the output to three related functions and files because Zheng~\emph{et al.}~\cite{zheng2025lvlms} observed that LLMs are prone to  hallucinate in free-form and longer outputs.  We used the Python script to obtain all the functions and files requested by LLMs. Note that we do not provide LLMs the entire project because of the context size limit of LLMs although LLMs can have large context size. We used the following prompt form for information collection LLMs, where \{bug\_report\} is the bug report and \{filename\} is filenames of source code files.

\begin{framed}
{\fontsize{7.5}{7.2}\selectfont
\setlength{\baselineskip}{7.2pt}
\begin{lstlisting}[
  aboveskip=0pt,
  belowskip=0pt,
  lineskip=0pt,
  breaklines=true,
  breakatwhitespace=true,
  breakindent=0pt,
  breakautoindent=false,
  columns=fullflexible
]
Given the bug report {bug_report}, can you tell me which methods and files would you look at based on the bug report. Please use the format: METHOD:1.[method] FILE:1.[FILE]. Please do not make up anything and just provide me the name of the methods and the name of the files without any explanation. Please only find three most important methods/files and please do not include () or any other information in the method name. The file should be in {filename}. Please be sure to reason the method name or file path based on the bug report and do not make up any method name as your job is to look at the bug report and find the related methods only. Note that you do not need to provide the file if you do not see the file. In this case, you can just say METHOD:1.[METHOD] FILE:1.NONE If you provide the file, please follow the provided file path. If you do not know the name of the method, please just give me the file name. In this case, you just need to give METHOD:1.NONE FILE:1.[FILE].
\end{lstlisting}
}
\end{framed}

\subsection{Static Analysis}
Static analysis is as shown in Area 2 in Figure~\ref{fig:overview}. We first used LLMs to decide which static analysis tasks to use. Then,  we used tools to generate static analysis results based on the decision that LLMs make. We used tools to generate the static analysis results because LLMs are susceptible to hallucinate on static analysis tasks~\cite{su2025code} and more complex source code~\cite{xie2025core}. LLMs are only responsible for reasoning which static analysis task to use and which paths to look at. We limited LLMs to two choices, i.e., callgraph generation and dataflow analysis. We limited LLMs to only two choices because these are two main strategies that human programmers use for bug localization~\cite{ko2004designing, smith2025human}. In addition, LLMs are required to provide the source and sink when dataflow analysis is chosen. We used Joern~\cite{joern} for dataflow analysis and Doxygen~\cite{DoxygenWebsite} for callgraph generation. We used Joern and Doxygen because these tools are used in several research papers~\cite{steenhoek2024dataflow, chen2023acer, csaszar2020interactive}. To reduce the length of the tokens, we used the same LLMs to find the most related call chains in the callgraph. This is because the size of callgraph could potentially become very large and not all of the call chains are related. We use the following prompt for deciding which static analysis to use, where \{bug\_report\} is the bug report from the user, \{codeblocks\} is the source code of the entry point, and \{function\_name\} is the name of the entry point functions.

\begin{framed}
{\fontsize{7.5}{7.2}\selectfont
\setlength{\baselineskip}{7.2pt}
\begin{lstlisting}[
  aboveskip=0pt,
  belowskip=0pt,
  lineskip=0pt,
  breaklines=true,
  breakatwhitespace=true,
  breakindent=0pt,
  breakautoindent=false,
  columns=fullflexible
]
Now, you can use call graph analysis, a method to analyze how the methods call each other and data flow analysis, a method to analyze how data flow in the problem. Given the bug report {bug_report} and the related methods {codeblocks}, could you tell me which method would you use to analyze the bugs? Please just tell me what you would use and be sure to refer to the context and not make up anything yourself. Please use the template: data flow analysis: source:[SOURCE] sink:[SINK] if you think data flow analysis is an appropriate method. Please note that source should be the name of the method in {function_name} and sink should be in a function that is inside the same method (if this is not the case, please suggest call graph) and do not provide any other information and the example is data flow analysis: source:[A] sink:[B] where A is the method name and B is the function name. otherwise: call graph analysis
\end{lstlisting}
}
\end{framed}

In addition, we use the following prompt for selecting the call chain, where \{callpath\} is the callgraph path from the tool.

\begin{framed}
{\fontsize{7.5}{7.2}\selectfont
\setlength{\baselineskip}{7.2pt}
\begin{lstlisting}[
  aboveskip=0pt,
  belowskip=0pt,
  lineskip=0pt,
  breaklines=true,
  breakatwhitespace=true,
  breakindent=0pt,
  breakautoindent=false,
  columns=fullflexible
]
Now, can you look at the call path {callpaths} and tell me which one do you need to locate the bugs based on the bug report {bug_report} and related methods {codeblocks}. Please give me the complete paths as shown in call path with the template path: 1.[CALL PATH] The between method sign is "<-" or "->". Please only tell me the paths that help to localize the bugs and the call chain should be in the provided paths and please only provide the call chain without any explanation. Please follow the call chain that I give you. Do not have () in the call chain.
\end{lstlisting}
}
\end{framed}

\subsection{Bug Reasoner}
The bug reasoner is as shown in Area 3 in Figure~\ref{fig:overview}. The input for the bug reasoner is the output from Area 1 and Area 2. The output of the bug reasoner is the reasoning steps for localizing bugs, the hypothesis, and whether there is any missing function. The bug reasoner starts with deciding which strategies to use for bug reasoning. The strategy includes forward reasoning, backward reasoning, and code comprehension~\cite{bohme2017where}. Forward reasoning is to follow the computational steps forward. Backward reasoning is to reason the bug backward from the unexpected point. Code comprehension is to comprehend the source code for bug reasoning. We guided LLMs with the previous static analysis results and three bug localization strategies to reduce hallucination and asked LLMs. The bug reasoner can also request more functions if needed. The conditions that need more information occur when the necessarily functions in the callgraph are not included. We used the following format as the prompt if callgraph analysis is used and the similar template for dataflow analysis, where \{callmethods\} is the functions in the callgraph and \{path\_to\_explore\} is the call chains.

\subsection{Language Models}
We used GPT-o4 as our base model with the cutoff on April 16, 2025 for bug localization. We used GPT-o4 as our models online because GPT-o4 represents one of the commercial models with strong reasoning skills although we also conducted the experiments with GPT-5 mini. In each area, we used separate LLMs to reduce the length of the tokens.

\begin{framed}
{\fontsize{7.5}{7.2}\selectfont
\setlength{\baselineskip}{7.2pt}
\begin{lstlisting}[
  aboveskip=0pt,
  belowskip=0pt,
  lineskip=0pt,
  breaklines=true,
  breakatwhitespace=true,
  breakindent=0pt,
  breakautoindent=false,
  columns=fullflexible
]
The strategies to localize the bugs are forward reasoning, backward reasoning, and code comprehension. Give you the related methods {codeblocks} and methods in call graph {callmethods}, and the call chain {path_to_explore}, could you use one of the strategies to reason the bugs and localize the bugs based on bug report {bug_report}? Please do not assume any information that is not provided, any information not in call chain and related methods, the code snippets not in the provided code, and the name of the method. You should only look at the provided method to reason the bugs. If the methods that you need are not in the call chain or related methods, you can request the methods. Please remember to choose either forward reasoning, backward reasoning or code comprehension to reason where the bugs localize and keep it consistence with the previous reasoning methods if you have and do not generate new code snippet or new call chain for the specific methods as your task is not to generate anything. Instead, your task is only to localize the bugs. Please use the template REASONING METHODS: [METHOD] REASONING STEPS: [STEPS] Hypothesis: [HYPOTHESIS] METHOD MISSING: [METHOD MISSING] (if you have multiple methods and please only provide method name and no other information is needed). Please do not make up any method name and please follow the call chain. Otherwise, use REASONING METHODS: [METHOD] REASONING STEPS: [STEPS] Hypothesis: [HYPOTHESIS]. Please do not make up anything outside the provided information, but you can request it if needed. The reasoning method should be the same as the previous one if you have already chose one and you should follow the call chain and please make your reasoning steps specific and do not make up any method name.
\end{lstlisting}
}
\end{framed}

\subsection{Web-based Interface}
\begin{figure}[h]
    \centering
    \vspace{-3mm}
    \includegraphics[width=0.85\linewidth]{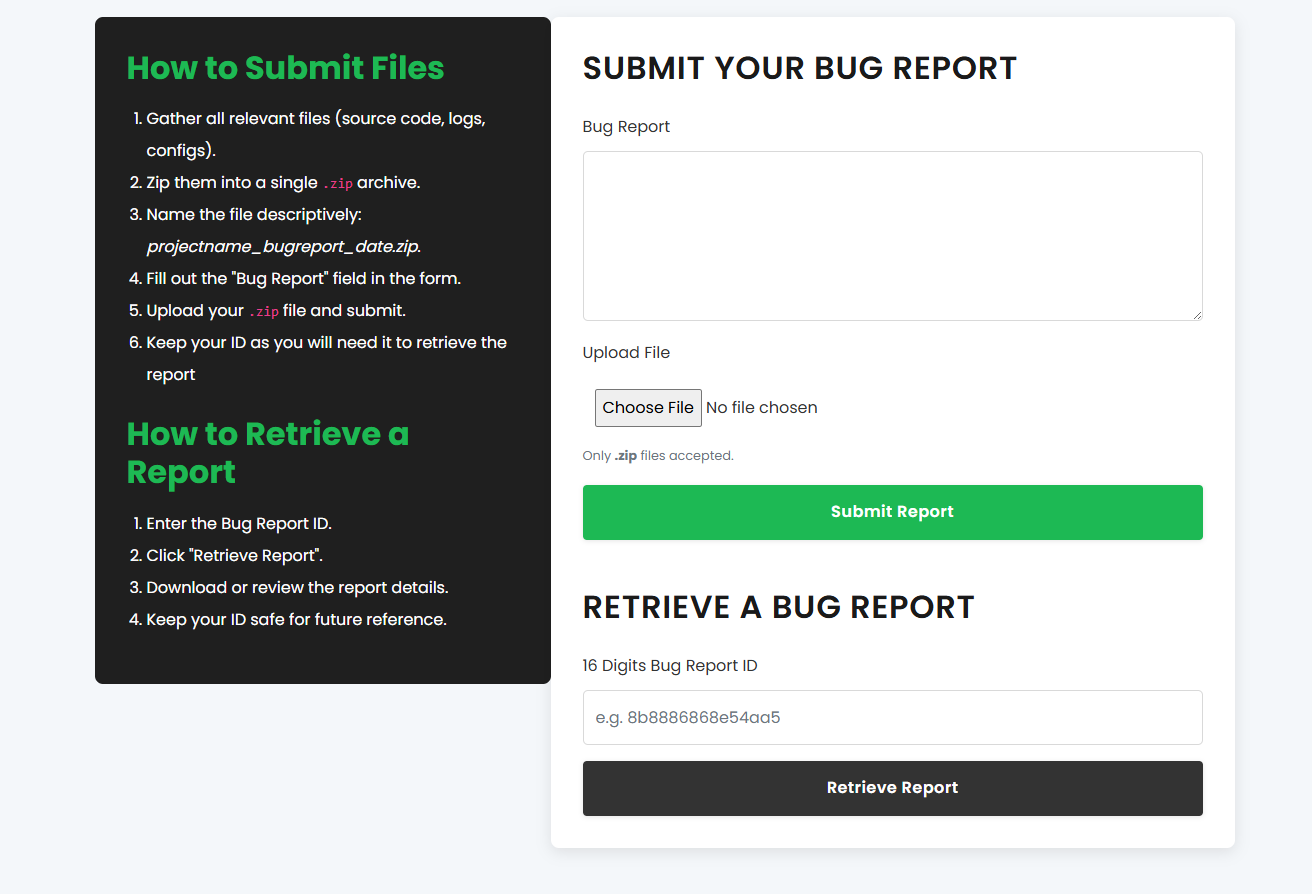}
    \caption{Web-based Interface for CMind}
    \label{fig:interface}
    %\vspace{-5mm}
\end{figure}

We developed a web-based interface for users to submit the bug report and source code as shown in Figure~\ref{fig:interface}. Users also need to use the same web-based interface to retrieve the results. The key idea is that users submit the necessary information. Then, the system will generate an ID number for that specific bug report and users. Users need to use this ID number to retrieve the bug localization results when it is done. CMind also allows users to download the results with the same ID number. The key requirements are as follows:
\begin{description}
\item \textbf{Bug report} A bug report should include as much detailed information as possible. For example, it could be the stack trace or the output from AddressSanitizer. It may be less accurate if you only provide the unexpected behaviors.  
\item \textbf{Source code} A source code package should include all files in the project because CMind uses those files for static analysis such as generating callgraph.
\end{description}

\section{Evaluation}
In this section, we discuss the dataset that we used for evaluation and the evaluation results with an example.

\subsection{Dataset}
We used the Heap dataset from Katzy~\emph{et al.}~\cite{katzy2025heap}. The Heap collects the dataset from the repositories with the licenses that forbid those companies to train LLMs with their code. The Heap also runs the deduplication to avoid duplication in the dataset. In addition to the Heap, we also used some bug reports from Redis with the report date on July, 2025. In total, we evaluated CMind with 20 bug reports. Note that most of the bug reports that we picked are with the clear stack track/AddressSanitizer or clear steps on how the bugs occur, so LLMs have clear steps to follow. We leave the one with only the symptom of bugs in our future works. 

\subsection{Preliminary Results}
We found the similar results for both GPT-5 mini and GPT-o4 although we found that GPT-5 usually has longer output compared with GPT-o4.  We show the results in Table~\ref{tab:result}. Specifically, we observed 75\% accuracy in GPT-o4 models and 80\% accuracy in GPT-5 mini models. Interestingly, we found that GPT-o4 and GPT-5 mini models fail in the same cases. For example, we found that LLMs usually go off-track when it does not have the clear stack trace/compiler error messages to follow. For example, in result12 in our repository, we found that the user only mentioned the steps to reproduce without providing any clues on where the unexpected behaviors occur. This causes output of LLMs to go off-track. This finding suggests that LLMs can be effective only with clear guidance or leash. Note that the problems are not trivial after providing stack trace/compiler error message because stack trace provides a hint for LLMs instead of an answer. This is also observed by Smith~\emph{et al.}~\cite{smith2025human} that some human programmers may still spend a significant amount of time localizing memory bugs even though some bugs may seem easy. We show an example in Section~\ref{sec:example} that this information provides a hint for LLMs to look at other similar functions for bug localization instead of an answer.

\begin{table}[h!]
\centering
\small
\caption{Result of Bug localization with GPTo4 and GPT5}
\begin{tabular}{|c|c|c|c|}
\hline
\textbf{Models} & \textbf{Number of reports} & \textbf{Correct} & \textbf{Incorrect} \\ \hline
o4-mini-2025-04-16 & 20 & 15 & 5 \\ \hline
gpt-5-mini-2025-08-07 & 20 & 16 & 4 \\ \hline
\end{tabular}

\label{tab:result}
\end{table}

\subsection{Example}
\label{sec:example}
In this section, we show an example that CMind could refer to other similar functions for bug localization. 

We found that the problems are not trivial after providing the stack trace because stack trace provides a hint for LLMs instead of an answer. Example 1 is the summarization of the hypothesis generated by CMind. The input includes the stack trace that shows where the bugs originate from. Overall, we found that CMind could refer to similar functions for bug localization. For example, in Step 3, CMind mentioned that other functions ``DisplayCaptureToolbar``, ``AudioCaptureToolbar``, and 
``WindowCaptureToolba`` have the conditions that check whether mod is NULL. On the other hand, ``ApplicationAudioCaptureToolbar`` does not have the condition that checks whether mod is NULL. Therefore, when mod is NULL, we the source code dereference a NULL pointer, which causes a crash. In addition, we also found that CMind also mentioned the reason that mod is NULL as shown in step 5. Overall, this result aligns with the findings by Smith~\emph{et al.}~\cite{smith2025human} that programmers visual attention focuses on several functions instead of individual functions, which suggests the effectiveness of our approach.

\begin{table}[h!]
\small
\vspace{-1mm}
\begin{tabular}{ll}
\textbf{Example} &  ID 7a3ed78200857e0\\ \hline
\end{tabular}
\begin{tabular}{l}
\end{tabular}

\begin{tabular}{p{8cm}}
Here’s a concise summary of the bug chain based solely on the provided 
information:

1. The helper function
bool obs\_module\_get\_locale\_string(const obs\_module\_t *mod, …)
assumes its first argument (`mod`) is non-NULL and does
`mod->get\_string(…)` without ever checking `mod` itself.

2. obs\_module\_get\_locale\_text calls obs\_module\_get\_locale\_string without 
adding any null-check around its `mod` parameter.

3. In all other toolbars (DisplayCaptureToolbar, AudioCaptureToolbar, 
WindowCaptureToolbar), after fetching a module via get\_os\_module they 
explicitly do
if (!mod) return;
thus avoiding any null dereference.

4. ApplicationAudioCaptureToolbar::Init, however, calls
obs\_get\_module("win-wasapi") → mod
then immediately calls
obs\_module\_get\_locale\_text(mod, …)
without checking whether `mod` is NULL.

5. On Linux, obs\_get\_module("win-wasapi") always returns NULL, so 
ApplicationAudioCaptureToolbar::Init ends up passing a NULL `mod` 
into obs\_module\_get\_locale\_text, which in turn passes it into 
obs\_module\_get\_locale\_string and dereferences mod->get\_string, causing 
the crash.

Hypothesis
ApplicationAudioCaptureToolbar::Init fails to guard against a NULL module 
before requesting localized text, leading to a NULL-pointer dereference 
inside obs\_module\_get\_locale\_string.

\end{tabular}
\vspace{-5mm}
\end{table}

\section{Conclusion}
This paper moves the state-of-the-art forward in three key ways:
\begin{enumerate}
    \item We developed our method for hypothesis generation based on a human study with programmers designed for C memory bug localization~\cite{smith2025human}. We showed that it is possible to develop an agent based on human behaviors.
    \item We evaluated our method on both GPTo4 and GPT5 with 20 bug reports. We found that GPTo4 reached 75\% accuracy and GPT-5 reached 80\% accuracy, which show that both models have similar results after we put the leash on LLMs although GPT5 has longer outputs compared with GPTo4.
    \item We found that LLMs usually go off-track when LLMs do not have clear instructions to follow or when we do not keep LLMs on a leash. This result aligns with findings from ~\cite{karpathy2024software} that LLMs have unlimited answers and can move off-track quickly. This implies a future development on agentic AI should focus more on limiting the output of LLMs.

\end{enumerate}

We release the source code and results in our Github repository~\url{https://github.com/apcl-research/CMIND} and make our tools and documents available in~\url{https://apclbuglocalizer.github.io/}. Users can use our online website to submit their bug report or use our Github code with Docker to build CMind in their local machines. 

\section{Future Works}
We divide our future works into four directions. First, we aim to integrate CMind into IDE/editors such as VS code to increase the usefulness of this tool. Second, we aim to increase the robustness of our approach. Currently, CMind could only localize bugs with clear stack trace. We aim to improve CMind by conducting experiments with more diverse bug reports and complete analysis, so it could localize bugs with implicit symptoms. Third, we aim to conduct the experiments with more models to improve the generalizability of our prompts. Finally, we aim to develop a model that can be run locally to protect the data privacy.
%In the future, we aim to have CMind localize memory bugs in Github develop an open-source model to protect privacy.
% provide users to adjudct how many functions
%robustness capbility
% more thorough analysis
%IDE integration
% generalibility of prompts

\section*{Acknowledgment}
This work is supported in part by NSF CCF-2211428. Any opinions, findings, and conclusions expressed herein are the authors and do not necessarily reflect those of the sponsors.

\bibliographystyle{ACM-Reference-Format}
\bibliography{main}
\end{document}